# Urdu News Article Recommendation Model using Natural Language Processing Techniques

___________________________________________________________________________

Syed Zain Abbas
MS Computer Science
Bahria University Islamabad

Dr. Arif ur Rahman
Associate Prof / HOD CS Department
Bahria University Islamabad

Abdul Basit Mughal
MS Data Science
Bahria University Islamabad

Syed Mujtaba Haider
MS Data Science
Bahria University Islamabad

___________________________________________________________________________

**Abstract:**

There are several online newspapers in urdu but for the users it is difficult to find the content they are looking for because these most of them contain irrelevant data and most users did not get what they want to retrieve. Our proposed framework will help to predict Urdu news in the interests of users and reduce the users' searching time for news. For this purpose, NLP techniques are used for pre-processing, and then TF-IDF with cosine similarity is used for gaining the highest similarity and recommended news on user preferences. Moreover, the BERT language model is also used for similarity, and by using the BERT model similarity increases as compared to TF-IDF so the approach works better with the BERT language model and recommends news to the user on their interest. The news is recommended when the similarity of the articles is above 60%.

___________________________________________________________________________

1. **Introduction**:

In this Modern time, there is a lot of information is about anything available on the internet mostly in the last few years. Today people in the world are astounded by the amount of data over and over it makes it difficult to choose somewhat from a huge number of preferences. Meanwhile, recommendation makes sense to recommend which thing should be shown to a user in which the user will have a phase to investigate and choose the wanted result [1]. Recommendation Systems (RS) are playing a vital role in current era of technology. In the year 2006, Netflix provides streaming movies and different TV series around the globe. Netflix announced a competition predict user ratings for movies, which is based on prior ratings that have no material relevant to a user or a movie. The Awards were founded on refining Netflix's algorithm. The team finally won the race and was capable to reach a 10% enhancement on Netflix's procedure, and they were bestowed with 1 million US $[2].

Usually, recommendation systems are composed of the following:

- **Users:** In the recommendation system people like to read a type of book or type of news article they want to read and for each user, the model can be concrete.
- **Items:** The things which the system is selected to recommend are called items. For every item, there are different belongings and features. For example, a writer of a news article stars in a film.

- **Preferences:** Preferences describe user like and dislike, for example, a user like a movie or reads an article, etc.

Recommendation Systems are essentially software implementation and procedures for information extraction and purifying that target to provide significant and operative item suggestions to the user. This prediction is also named as a recommendation related to many results like which news article is recommended for a reader, which type of product or a thing to purchase, and what type of song is recommended for a user to listen to. Recommendations systems are generally divided into two type personalized news recommendations and non-personalized news recommendations. In the personalized news recommendation, all users or groups of users have dissimilar recommended content. Whereas in non-personalized news recommendations all users or groups of users have similar recommended content.

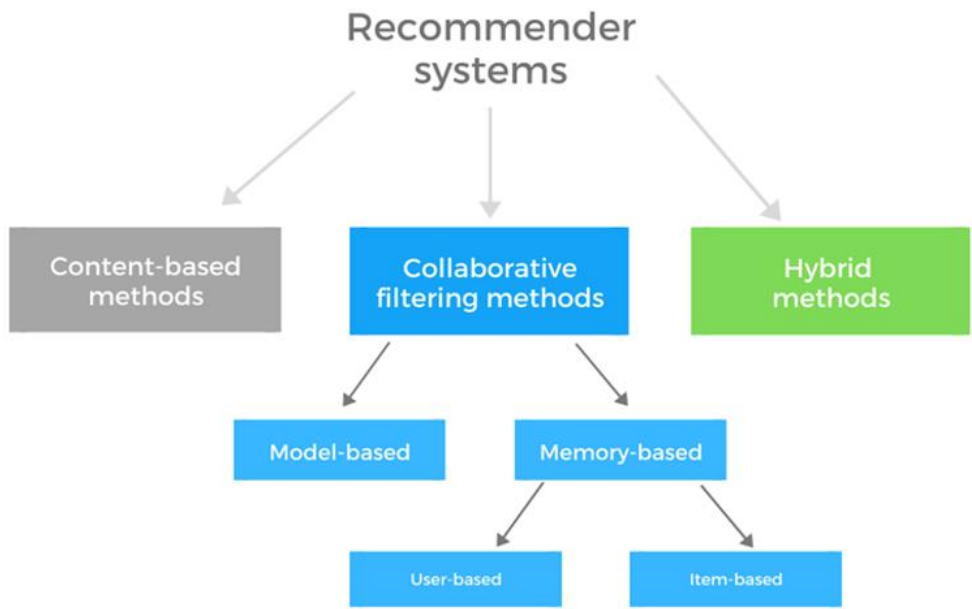

*Figure 1: Recommendations Systems.*

### 1.1 Non-personalized systems:

It includes statistics and by using exterior data from the public like which products people like the most and which thing is selling at trending. It also includes how much people like this thing (item). Item in terms of different things like movie, song, dish. For example, how much the population like a restaurant is converted into a list, and this list has used the suggestion of the best restaurant in the city [3].

### 1.2 Content-based filtering system:

In the Content-Based recommend system users like items and things in which he/she is interested and upon user interest, the model tries to build on these preferences [4]. An example from the sphere of movies. Assume someone likes science fiction movies, and action and fantasy movies, but did not like romantic movies. Eventually, the algorithm can collect records, and based on this the system investigates that the user has affirmative marks in categories like science fiction, fantasy, and action, but a lesser score for romantic movies. The algorithm also finds out that there are some actors whom the user likes or loathes. Our approach fits in this type of algorithm as we are trying to build a model on user feedback from news articles. Figure 1.2 delivers a virtuous overview of Content-Based Filtering.

### 1.3 Collaborative-based filtering approach:

This approach is based on the history information of the users and this approach has been broadly used in different applications. The very first system which produce automated recommendations was the Grouplens system [5]. The said system provided a personalized recommendation on Usenet posting. In Collaborative-based filtering, we tend to search out parallel users and suggest what similar users like. During this type of recommendation system, we don't use the options of the item to suggest it, rather we tend to classify the users into clusters of comparable sorts, and suggest every user in step with the preference of its cluster.

### 1.4 Hybrid recommended systems:

In this recommended system there is a combination of two or more recommendation approaches together to achieve improved performance these are utilized to gain the maximum of the utilization of some approaches and avoid the minimize drawbacks of some approaches [6].

## 2. Literature Review:

In this paper, the approach used for recommendation is that the given approach provides different news links to a user on their interest instead of providing news articles. The news URLs download from the GDELT. In the approach different news are categorized according to different news sources that which source is best of which news articles [7].

In [8], a well-personalized news recommendation technique is used. For this purpose, a user profile is built with three different characteristics: based on keywords, news topic dispensation, and label entities. The user profiling method named BAP which is calculated the user preferences for the news articles based on the user behavior and news reputation. Furthermore, the profile of the user can be built more accurately with this method. A web-based approach named Newsday that delivers news the aim is to deliver news in the Thai language, BreakIterator9 is used to tokenize Thai text. Noise words are eliminated with Lexitron10 by cross-checking each token. First data is extracted from different website in a raw form then manipulated and store in a MySQL database, the web application works in two ways first by using this web application user first create an account and then select a category, so the application recommend news according to user preferences and second it stores the history of a user and deliver news according to user preference. The accuracy of both ways is 73% and 84% respectively [9].

### 2.1 Approaches using Machine Learning:

In the past couple of years, ML is very useful for solving-knowledge problems in the field of computer science. In the paper, they proposed a words-clustering system to manipulate the given script and bunch all arguments with like aspects. The method contains two phases, the first processing and the other is the clustering phase. In the prior processing, they tokenize the input text and control the word sequence within the text by giving a number to each word. After this word-class identifies by the lexical search i.e a word belongs to which class and whether the class belongs to a noun or pronoun [10].

A framework was proposed to display news articles using user preferences, the system will store the user's interest on the bases of likes and dislikes, and the time utilized by the user for reading a news article and will update user profiles, the news extract from different reliable sources using NLTK library to extract a keyword from title and description and store, the store news article are then classified into various types using SVM and Naïve Bayesian classifiers and they use k-means unsupervised clustering algorithm for labeling each cluster for subcategories of news [11].

A multi-agent structure was proposed for positioning news articles on the premise of users' interests raised from social mass media. To do so, they take incontestable the connection between users' social media favorites and news categories: they need deep-mined categories from social media, premeditated with general news classes. The developed resolution provides twenty-seventh improved results than gift news websites recommendation. Supported these ideas, they need industrialized associate mechanical man application and performed a pilot study. The results show higher satisfaction levels for users once looking out for news articles through the planned system [12].

### 2.2 Approaches using Deep Learning:

In this paper, the multilingual information is extracted using deep learning which supports three languages English Hindi and Marathi. The first user gives an input sentence, NLP is used to process this sentence, and word to vector is used to convert the words into numeric values, for the word to vector CBROW is used to convert word to numeric values than three-layer deep learning model take this input for the training process. Finally, RNN is used to obtain similar words according to the user query and it is done by topic modeling. If the word does not match the RNN will obtain based on training. Hence finally matching records are retrieved using the highest score Cosine similarity by using deep learning the accuracy score is 91% and without deep learning the accuracy is 70%[13].

In this work, they use a deep neural network consideration to challenge the problem of news recommendation. This model is a hybrid of user-item collaborative filtering which uses the content of the items as well. The main aspect of user-item-based collaborative filtering is to detect the communication between a user and item features. Matrix factorization is one of the most communal methods for recognizing this communication. They present a deep neural model, where a non-linear plotting of users and item countryside are learned first. For knowledge of non-line plotting for the handlers, we use a consideration-based recurrent layer in a blend with fully connected layers. For knowledge, the plotting for the kinds of stuff they custom only fully connected layers [14].

### 3. Methodology:

A recommendation system is a very useful way to solve information overload. Recommends interesting news to users depending on the user's behavioral data and interest. The recommendation system can calculate similarities by learning the user's interests and preferences and ultimately helping users find their information needs. The purpose is to make an Urdu NR approach that recommends Urdu news depending on user interest. In this approach, the user read the news article, and the system store user's read article Id furthermore it compares the read news article with the entire dataset which contains different news articles comparing similarity is counted based on the user's read article and the entire dataset news articles and highest similarity news articles then selected for recommendation to a user. First the approach read dataset the dataset is in CVS format after reading the dataset the rows with missing values is drop and now refine dataset is use for further steps include pre-processing in pre-processing the text first we remove the white spaces from the text white spaces are the spaces between different words after removing white spaces the text is in normal form the next step is to remove the punctuation removal in this step all the punctuation marks e.g. ",,:" are remove from the text this punctuation marks are meaningless and not in use, after remove the punctuation marks the text is cleans from the punctuation marks, after this stage we remove accents from the text accents means Zabar, Zair, Paish all this accents are remove from the text, so after this stage text is free from this accents, after removing accents the next procedure is to remove the URLs in the text URLs contains different website links in the text which are not useful so we remove URLs from the text, after removing URLs the next step is to remove emails from the news articles text, if the text contains any emails so this process remove emails from the text, after removing the emails the next step is to remove the phone numbers, if the text contains phone number then phone number is remove from the text, after removing phone numbers we also remove numbers from the text, articles contains any number will be remove from the article, after this currency symbols are replace e.g. $ is replace with USD, the next step is to remove the English alphabets from the text, all this pre-processing technique using Urdu Hack library, now text is ready or tokenization, in tokenization we split the text in to words, then stop words remove from the tokenize text, stop words are those words which are commonly use in all the articles and does not have some meaning. Urdu hack and the spacy library contain Urdu stops words, so we compare our news articles with the stops word provided by these libraries after all these steps a simple and refined news article text is used for similarity this processing is applied to useful articles, and as well as all entire data set. after this we use TF-IDF with cosine similarity and Bert language model for similarity count and gaining highest similarity if the similarly above 0.60% we recommend the news basis on user read articles.

Our proposed method recommends news according to user preferences and develops an understanding that how useful features are extracted using multiple NLP techniques for developing a news recommendation approach.

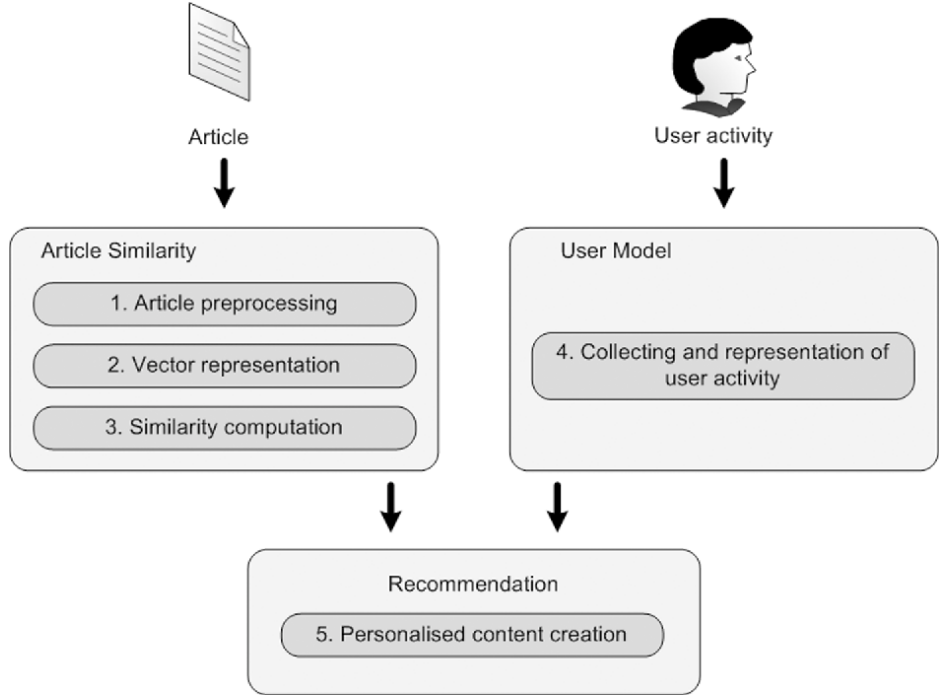

*Figure 2: Workflow of the proposed methodology.*

### 3.1 Dataset:

The first step is to select a source that can use for building the data set. A Comprehensive News Data set for Urdu Text Processing the data set is available for solving many NLP, Machine/Deep Learning tasks: text processing, classification, summarizing, Named Entity Recognition, Topic Modelling, and Text Generation. This data set includes Urdu news articles related to four categories: Business & Economics, Science & Technology, Entertainment, and Sports in each category multiple news articles are placed. Table 3.1 shows the details of the data set, Business, and Economics category contains 300 articles, Science and Technology contain 330 articles, Entertainment contains 300 articles and sports contains 230 articles.

| Category | No of articles |
| --- | --- |
| Business and Economics | 300 |
| Science and Technology | 330 |
| Entertainment | 300 |
| Sports | 230 |

Dataset Details

We show one news headline and one news article from every category after converting it from Unicode text to inpage Urdu text.

**Headline of Business news article**

عالمی بینک عسکریت پسندی سے متاثرہ خاندانوں کی معاونت کرے گا

*Figure 3: Headline of the Business news article.*

The above figure shows the headline text of the Business & Economics news article the shown copy from the dataset and converted it to inpage Urdu text using Unicode.

*Figure 4: News text of business news article*

## 4. Evaluation:

The Result was evaluated by applying NLP techniques to the dataset tasks of NLP including Tokenization, removal of stop words, and so on and different similarity techniques are applied to the dataset to achieve the highest similarity and recommend news for the user on their preferences. The below stages are done by the present technique:

### 4.1 Pre-processing of news articles:

In this phase, different NLP methods like (tokenization, stop word removal, punctuation remover, whitespaces remover, URLs remover, phone number remover, etc) are applied to the source News.

### 4.2 Particular collation phase:

This phase matches the User's read news with default news and calculates the matching score using TF-IDF and BERT language model.

Two approaches are used for the evaluation of the proposed approach which is as follows:

### 4.3 TF-IDF:

TF-IDF can be broken down into two corridors TF (Term Frequency) and IDF (Inverse Document Frequency). It is used in the fields of information reclamation and machine literacy. It's a qualitative measure that assesses how much related a word is to a document and in a collection of documents. The Process is done by proliferating 02 matrices how numerous times a word shows in a document and the other document frequency of the word across a group of documents. It works by adding proportionately to the no of intervals a word shows in a document but is counterpoised by the no of documents that contain a word. So, words that are common in every document, similar as this, what, and if, rank low indeed though they may appear numerous times since they don't mean important to that document. TF-IDF for a word in an article is evaluated when two dissimilar metrics are cross multiple: The TF of a word in an article. There are a few ways of manipulating this frequency, with the greenest being a fresh count of instances a word seems in a document. Then, there are ways to correct the frequency, by the length of a document, or by the raw frequency of the most frequent word in a document. The inverse document frequency of the word through a set of articles. This relieves, in what way common or rare a word is in the entire document set. The nearer it is to 0, the more related a word is. This metric can be calculated by captivating the total number of documents, dividing it by the number of documents that enclose a word, and calculating the logarithm. So, if the word is precise mutual and seems in many documents, this number will take 0. Otherwise, it will take 1.

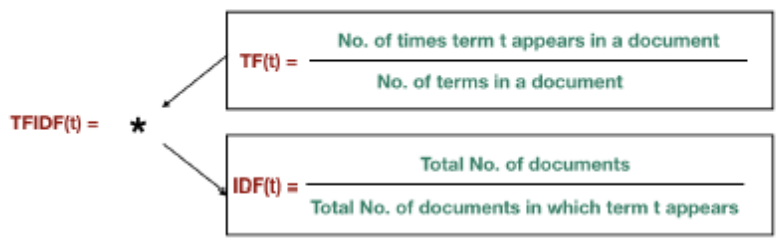

*Figure 5: TF-IDF*

### 4.4 BERT Language Model:

BERT language model is an open-source machine learning structure for (NLP). BERT is aimed to help computers understand the meaning of ambiguous language in the text by using adjacent text to establish context. Transformers were first presented by Google in 2017. At the time of their starter, language models mostly used (RNN) and (CNN) to handle NLP problems. Even if these models are capable, the Transformer thinks about a major development because it doesn't require the continuance of data to be treated in any fixed order, whereas RNNs and CNNs do. As Transformers can process data in any order, they enable the preparation of larger amounts of data than ever was possible before their presence. This, in turn, helped the formation of pre-trained models like BERT, which was trained on huge amounts of language data earlier to its announcement. In its investigation stages, the framework accomplished revolutionary results in 11 usual language understanding tasks, including sentiment analysis, semantic role labeling, sentence classification, and the disambiguation of polysemous words, or words with multiple meanings. Finishing these tasks recognized BERT from earlier language models such as word2vec and GloVe, which are inadequate when understanding context and polysemous words. BERT capably reports ambiguity, which is the utmost task to natural language understanding according to research scientists in the field. It is capable of analyzing language comparatively like what humans think.

### 5  Experiment Setup and Results:

This section explains present details of work that how we apply pre-processing on data set and after this step we apply different similarity techniques for comparing news that user read with other news in the data set after that high similar news recommend to the user on the basis of their interest. Experiment setup was carried out using Google Colab in which we read the data set which is in CVS format. The data set contains different columns named as ID,

Headline, News Text, Category, and News Length. In the Data set, we select relevant columns and remove rows with missing values.

| News Text | Category | News length |
|---|---|---|
| اسلام آباد عالمی بینک خیبرپختونخوا کے قبائلی اص... | Business & Economics | 1854 |
| اسلام آباد فیڈرل بورڈ ریوینو ایف بی آر نے دسمبر کی... | Business & Economics | 2016 |
| اسلام آباد بورڈ آف انوسٹمنٹ بی او آئی کے چیئرمین ع... | Business & Economics | 2195 |
| اسلام آباد پاکستان میں ماہ نومبر میں مسلسل تیس... | Business & Economics | 2349 |
| اسلام آباد نیشنل ٹرانسمیشن اینڈ ڈسپیچ کمپنی این... | Business & Economics | 2655 |

*Figure 6: Dataset*

After that, we apply NLP techniques to the data set for prepossessing, tasks of NLP include Tokenization, removal of stop words, and so on, and different similarity techniques are applied to the data set to achieve the highest similarity and recommend news for users on their preferences. In the first phase we apply Pre-processing and second phase we compare different similar methods.

### 5.1 Pre-processing Techniques:

In article pre-processing, we perform tokenization, stop words removal, punctuation removal, etc on the news of user read articles and source news for this purpose we use Spacy library and Urdu Hack. the spacy library is used for Natural language processing in Python. It's engineered on the terribly latest analysis and was designed from day one to be employed in the real product. Spacy comes with pre-trained pipelines and presently supports tokenization and coaching for 60+ languages. It options progressive speed and neural network models for tagging, parsing, named entity recognition, text classification, and a lot of, multi-task learning with pre-trained transformers like BERT.

### 5.2 Normalize the text from white spaces:

To normalize some text, it will return a string with normalized characters both single and combined, proper spaces after digits, in this white space from the text is replaced/removed**.** For example, if a text is given which has spaced the technique removes white spaces from the text shown in the below figure.

```
Text with white spaces
عراق اور شام    اعلان کیا  بے  دونوں    جلد اپنے      گے؟
Text after removing white spaces
'عراق اور شام اعلان کیا بے دونوں جلد اپنے گے؟'
```

*Figure 7: Wide Spaces remove*

### 5.3 Punctuation Removal:

Punctuations are an effective part of written speech however, punctuations are worthless. In article pre-processing, punctuation did not have any meaning, for some reason during the process, we eliminate all punctuation marks in news texts. For example, in a text given which punctuation marks like '?' the technique removes the punctuation marks from the text which are shown below figure.

Text with punctuation
عراق ؟اور شام اعلان کیا؟ بے دونوں جلد اپنے گے؟
Text after removing punctuation
'عراق اور شام اعلان کیا بے دونوں جلد اپنے گے'

*Figure 8: Removal of Punctuations*

**5.4 Remove Accents:**

  Remove accents from any accented Unicode characters in the text string, either by transforming them into ASCII equivalents or removing them entirely. For example, if a text is given which contains accents the technique removes all the accents from the text which is shown in the below figure.

Text with accents
عدالتؑ عظمیٰ پاکستان
Text after removing accents
'عدالت عظمی پاکستان'

*Figure 9: Removal of Accents*

**5.5 Replace URLs:**

 Remove all URLs from the news text. For example, if a text is given which contains URLs the process removes all the URLs from the text which are shown in the below figure.

Text with url
فیصد www.gmail.com 20
Text after removing url
'فیصد 20'

*Figure 10: Removal of url*

**5.6 Replace Emails:**

  In this step, all the emails from the text are removed. For example, if a given text contains emails in this step all the emails from the text is removed which are shown in the figure below.

```
Text with email
20 gunner@gmail.com فیصد
Text after removing email
'20    فیصد'
```

*Figure 11 : Removal of Email*

**5.7 Remove numbers**:

    In this step, all the number in the text is removed. For example, if a text contains numbers the process is to remove all the numbers from the text which is shown in the below figure.

```
Text with Number
فیصد 20
Text after removing number
' فیصد '
```

*Figure 12 : Removal of numbers*

**5.8 Remove Phone Number:**

    In this step, all the phone number in the text is removed. For example, if a text contains a phone number the process is to remove all the phone numbers from the text which are shown in the below figure.

```
Text with Phone Number
یعنی لائن آف کنٹرول پر فائربندی کا معاہدہ 555-123-4567 میں ہوا تھا۔
Text after removing Phone Number
'یعنی لائن آف کنٹرول پر فائربندی کا معاہدہ   میں ہوا تھا'
```

*Figure 13 : Removal of Phone Number*

**5.9 Replace Currency symbols:**

    In this step all currency symbols from the text are removed. For example, a given text contains currency symbols that are removed from the text are shown in the below figure.

```
Text with currency Symbol
یعنی لائن آف کنٹرول پر فائربندی کا معاہدہ 2003 میں ہوا $33 تھا۔
Text after replace currency symbol with USD
'تھا۔ USDیعنی لائن آف کنٹرول پر فائربندی کا معاہدہ 2003 میں ہوا 33 '
```

*Figure 14: Replace Currency symbols*

**5.10 Remove English Alphabets**:

In this step, all the English alphabets are removed from the text. For example, if a given text contains a different English alphabet the current step removes all the English alphabet from the text which is shown below, figure.

*Figure 15: Replace the English alphabet.*

**5.11 Tokenization:**

Tokenization is the procedure of dividing input text into single words, tokens, and any additional meaningful word units. In this stage, we apply tokenization to Urdu text. Although tokenization in Urdu text processing is a, very difficult task with the help of Spacy and Urdu hack library Tokenization is performed, In these libraries, the text is tokenized based on sentences and words. Tokenization is shown in the below figure. This shows that a given text tokenizes into words after copying a Unicode text and covert into page Urdu text for a clear view.

*Figure 16: Tokenization*

**5.12 Stop words remove**:

Stop words are a major part of any speech, however, throughout the assessment of documents, they're not measured as vital. In any knowledge set, the frequency of stop words continually is high, thus it's necessary to get rid, of these stop words and minimize the document size. This method will increase the effectiveness and potency. The above-said libraries contain Urdu stop words that are removed throughout the method the Urdu stop words once changing from Unicode text to inpage.

*Figure 17: Urdu Stops Words in Spacy Library*

When we apply stop remover technique to a text he length of a text also reduces after apply stop words remover from the text are shown in the below figure.

```
length of token
14
length of token after stopword remove
11
```

*Figure 18: Text after applying stop word remove*

## 5.13 Comparison:

In this portion dissimilar similarity measurements to calculate similarity among Users view/read news articles and other news Articles. For this different similarity, methods are utilized which are as follows:

First, we apply TF-IDF to the news articles and Generate the Tf-Idf matrix model for the entire corpus then after that we apply TF-IDF to Generate the TF-IDF matrix model for reading articles TF-IDF is a method that calculates how related a word in a document or in a group of documents. This is possible by multiplying two metrics: how many times a word occurs in a document, and the inverse document frequency of the word across a set of documents). Then we use Cosine similarity to compare users who read news articles and the entire corpus. For example, a user read news articles on index 5 and we compare the similarity of that index with the index starting from 0 to 100 and allow only 10 recommendation news to be shown.

| Read Articles | Other Articles | Cosine Similarity Score using TF-IDF |
|---|---|---|
| Read News article on Index 5 | News Article at index 0 to 9 | 0.49 |
| | | 0.57 |
| | | 0.35 |
| | | 0.65 |
| | | 0.44 |
| | | 0.54 |
| | | 0.46 |
| | | 0.29 |

Figure 19: Similarity using TF-IDF cosine

| Read Articles | Other Articles | Cosine Similarity Score using Bert |
|---|---|---|
| Read News article on Index 5 | News Article at index 0 to 9 | 0.69 |
| | | 0.77 |
| | | 0.55 |
| | | 0.78 |
| | | 0.34 |
| | | 0.64 |
| | | 0.66 |
| | | 0.46 |

Figure 20: Similarity using Bert Language model

When applying Cosine Similarity using Bert Language Model similarity increase instead of cosine similarity using TF-IDF Model.

| Measure | Value | Derivations |
|---|---|---|
| Sensitivity | 0.8333 | TPR = TP / (TP + FN) |
| Specificity | 0.5 | SPC = TN / (FP + TN) |
| Precision | 0.7143 | PPV = TP / (TP + FP) |
| Negative Predictive Value | 0.6667 | NPV = TN / (TN + FN) |
| False Positive Rate | 0.5 | FPR = FP / (FP + TN) |
| False Discovery Rate | 0.2857 | FDR = FP / (FP + TP) |
| False Negative Rate | 0.1667 | FNR = FN / (FN + TP) |
| Accuracy | 0.700 | ACC = (TP + TN) / (P + N) |
| F1 Score | 0.7692 | F1 = 2TP / (2TP + FP + FN) |
| Matthews Correlation Co-efficient | 0.3563 | TP*TN - FP*FN / sqrt((TP+FP)*(TP+FN)*(TN+FP)*(TN+FN)) |

Figure 21: Performance metrics

6. Conclusion:

We have shown how news articles can be used for news recommendations to a user-on-user interest. Our main goal is to recommend Urdu news articles. To present this approach stage by stage that is how our approach developed.

First, we define the structure of the problem. Then, pre-process the data set and remove the rows which have missing values. After that, we extracted features from the data set by applying pre-processing to the data and normalizing the text by removing white spaces, numbers, phone numbers, English alphabets, URLs, emails, etc and we apply tokenization to a text and then we remove the stop words from the text now the text is in normalizing form. We used the extracted features from the text to recommend the news article for the user on their interest and preferences by using TF-IDF with Cosine similarity and Bert language Model. However, Cosine similarity with the BERT model has the highest similarity then TF-IDF. The news recommends according to the similarity above 60% between articles.